\documentclass[12pt,a4paper]{article}
\usepackage[english]{babel}
\usepackage{fullpage}
\usepackage{epsfig}
\usepackage{amssymb}
\usepackage{amsmath}
\usepackage{amstext}
\usepackage{cite}
\usepackage{indentfirst}
\usepackage{afterpage}

\title{Structure of the Circumbinary Envelope Around a Young Binary System}

\author{
P. V. Kaigorodov, D. V. Bisikalo, A. M. Fateeva, and A. Yu. Sytov
}

\begin{document}

\maketitle

\abstract{ The structure of a circumstellar envelope around a
young binary T Tauri star is considered. The supersonic orbital
motion of the system components in the envelope gas leads to the
formation of bow shocks around the star. Two- and
three-dimensional numerical modeling indicates an important role
of these shocks in the formation of the structure of the
circumbinary envelope. In particular, for systems with circular
orbits, the size of the central region of the envelope that is not
filled with matter (the gap) is essentially determined by the
parameters of the bow shocks. These modeling results are supported
by comparisons of the obtained estimates for the gap parameters
with observations. }

\section{Inroduction}

In an early stage of their evolution, close-binary systems pass
through a stage of accreting matter from a common envelope, which
is essentially a remnant of the protostellar cloud. Of all stars
located in this stage, the best studied are binary T Tauri stars
-- young stars that are just entering the main sequence.
Observations of these stars in the radio and infrared indicate
that their envelopes consist of gas ($\sim 99\%$) and dust ($\sim
1\%$). The gaseous component consists of molecular hydrogen
$\mathrm{H}_2$ and $\mathrm{He}$, together with a number of
heavier molecules -- $\mathrm{O}_2$, $\mathrm{CO}$,
$\mathrm{CO}_2$, $\mathrm{H}_2\mathrm{O}$ and others. These
envelopes have a disk-like form; their thickness grows with
distance from the center of mass of the system, and the velocities
of matter in the envelope display Keplerian distributions. The
presence of a region in the central part of the disk that is free
from matter -- so-called gap -- is also indicated by the
observations, at least for some systems.

\begin{figure}[ht]
\begin{center}
\epsfig{file=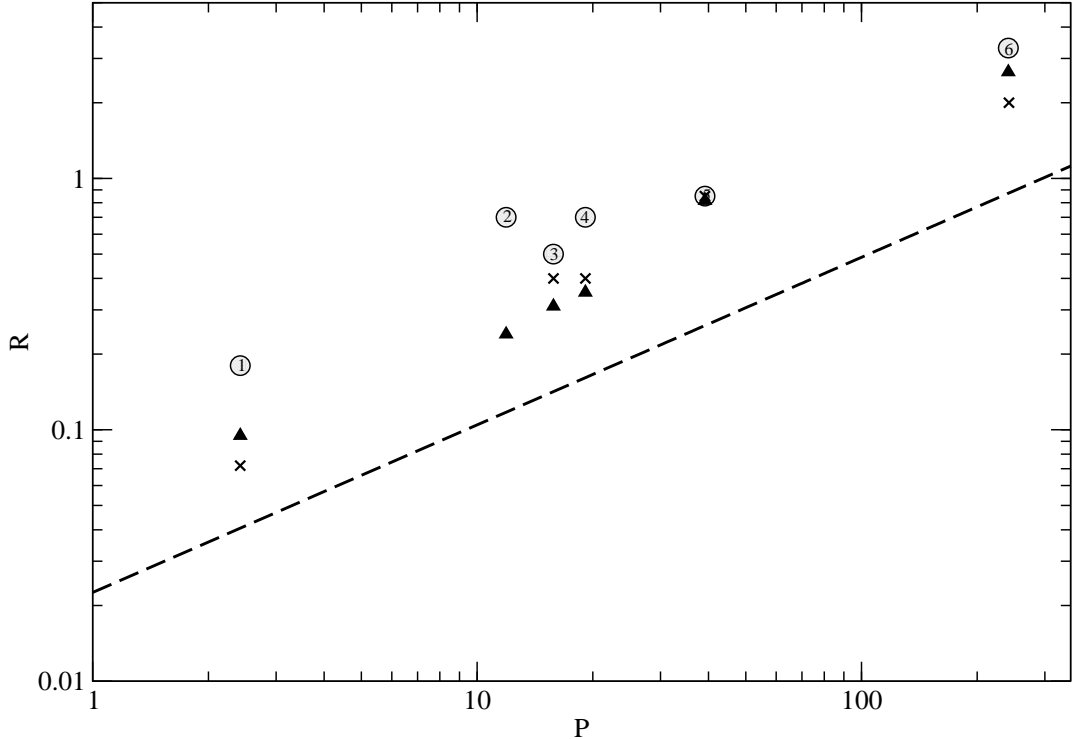, width=14cm}
\end{center}
\caption{ Dependence of the gap radius R for T Tauri stars on
their orbital periods P. The circles with numbers show the
observed radii for various systems, identified according to their
numbers in the table(~\ref{tbl1}). The "$\times$" show estimates
of gap radii corresponding to the location of resonances
~\cite{Artymowicz-Lubow:94}. The dashed line shows the minimum
possible gap radius, corresponding to the distance from the center
of mass to the outer radius of the accretion disk of the primary
component. The triangles show the gap radii derived from the
results of gas-dynamical modeling.}\label{modGap}
\end{figure}

The formation of gaps around binary stars was first studied by
Artymowicz and Lubow
~\cite{Artymowicz-Lubow:94,Artymowicz-Lubow:96}, who also carried
out numerical simulations of the accretion of matter from a
protoplanetary disk onto the binary. They concluded that the
leading role in the formation of inner gaps in the protoplanetary
disks of binary stars was played by Lindblad resonances. In spite
of the obvious influence of resonances on the structure of the
circumbinary envelope, it follows from ~\ref{modGap} and the table
that the theoretical sizes of such gaps estimated using the method
of ~\cite{Artymowicz-Lubow:94} lie systematically below the
observed values.

\begin{table}[ht]
{\small
\begin{center}
\begin{tabular}{cccccccccc}
N & Star & $R_{\text{obs}}$ (AU) & $R_{\text{res}}$ (AU) &
$R_{\text{mod}}$ (AU) & P (days) & $A$ (AU) & q & e & References\\
\hline
\hline\\
1 & V4046 Sgr & 0.18 & 0.072 & 0.0948 & 2.42  & 0.04  & 0.8  &
$\sim 0.01$ &
\cite{Jensen-Mathieu:97,Stempels-Gahm:2004}\\
2 & GM Aur    & 0.7  &       & 0.24   & 11.9  & 0.1   & 0.42 &
&
\cite{Marsh-Mahoney:92,Bouvier-et-al:95}\\
3 & DQ Tau    & 0.5  & 0.4   & 0.31   & 15.8  & 0.135 & 1    &
0.56 &
\cite{Najita-et-al:2003,Boden-et-al:2009}\\
4 & UZ Tau E  & 0.7  & 0.4   & 0.3519 & 19.13 & 0.153 & 0.3  &
0.33 &
\cite{Najita-et-al:2003,Jensen-et-al:2007}\\
5 & Roxs 42C  & 0.85 & 0.85  & 0.816  & 39.15 & 0.3   & 0.91 &
0.48 &
\cite{Jensen-Mathieu:97}\\
6 & GW Ori    & 3.3  & 2.0   & 2.65   & 241.9 & 1.13  & 0.32 &
0.04 &
\cite{Gullbring-et-al:2000,Mathieu-et-al:91,Mathieu:94}\\[3mm]
\hline
\end{tabular}
\end{center}
}\caption{ Gap sizes for various binary T Tauri stars
($R_{\text{obs}}$, $R_{\text{res}}$, and $R_{\text{mod}}$ are the
gap radii that are derived from observations, the location of
resonances, and gas-dynamical modeling; A is the component
separation, P the period, q the component mass ratio, and e the
eccentricity) }\label{tbl1}
\end{table}

One possible explanation for the systematically larger sizes of
the observationally derived gaps is failure to include certain
gas-dynamical effects in earlier studies of their formation. The
SPH method was used to numerically model the gas dynamics in
~\cite{Artymowicz-Lubow:94,Artymowicz-Lubow:96}, with a very small
number of particles, due to the limited power of computers
available at that time. As follows from the figures presented in
~\cite{Artymowicz-Lubow:94,Artymowicz-Lubow:96,Bate-Bonnell:97},
SPH particles are virtually absent from the inner region of the
envelope, making it impossible to adequately model the gas
dynamics of the matter in this region. However, gas-dynamical
computations carried out using grid methods
~\cite{Ochi-et-al:2005,Hanawa-et-al:2010} show the formation of a
complex flow pattern in the inner regions of the envelope,
including the accretion disks around the stars and a system of
shocks. Among the most prominent structural elements of the flow
are bow shocks formed due to the supersonic orbital motion of the
components of the system and their surrounding accretion disks
through the gaseous envelope.

The goal of the current paper is to study the influence of these
bow shocks on the flow structure in the inner regions of the
circumstellar envelope and, in particular, on the formation of
gaps in the envelope.

\section{The numerical model}

Our model of a binary star includes the system components and the
circumstellar gaseous disk, and is described by the masses of the
stars $M_1$ and $M_2$, their radii $R_1$ and $R_2$, the distance
between the component centers of mass $A$, the outer radius of the
circumstellar disk  $R_{ext}$, the equatorial density of the disk
$\rho_{disk}$, and the temperature of the disk $T_{disk}$.

At the initial time, the spatial region was filled with gas with
density $\rho_{disk}$ and temperature $T_{disk}$. Free-inflow
conditions were specified at the surfaces of the components.
Constant density and a constant velocity corresponding to the
Keplerian value at the given distance from the center of mass of
the system were specified at the outer boundary of the region. The
region inside the stars and outside the computational region,
$r>R_{ext}$, were excluded from the computations.

The modeling was carried out in a rotating coordinate system tied
to the binary star. We described the gas flows in the system using
a system of Euler equations for gravitational, adiabatic gas
dynamics, closing them with the equation of state of an ideal gas.
The temperature in the solution was held constant and equal to the
initial temperature of the initial disk, $2656 K$. The radius of
the stars was taken to be $1 R_\odot$ in all computations.

We solved the system of equations numerically using a
Roe–Osher–Einfeldt finite-difference scheme similar to those in
~\cite{AZ:2000:Bisikalo-et-al,BinaryStars:2002,AZ:2003:CoolDiscs}.
All the two-dimensional computations were performed in a
$12A\times 12A$ computational region on a uniform grid with
$2200\times 2200$ cells. The three-dimensional computations were
performed in a $12A\times 12A \times 1.5A$ computational region on
a non-uniform grid with $480\times 480 \times 224$ cells. The grid
used for the three-dimensional computations varied such that the
resolution between the components was no worse than in the
two-dimensional solutions.

\section{Results of the numerical simulations}

\begin{figure}[ht]
\begin{center}
\epsfig{file=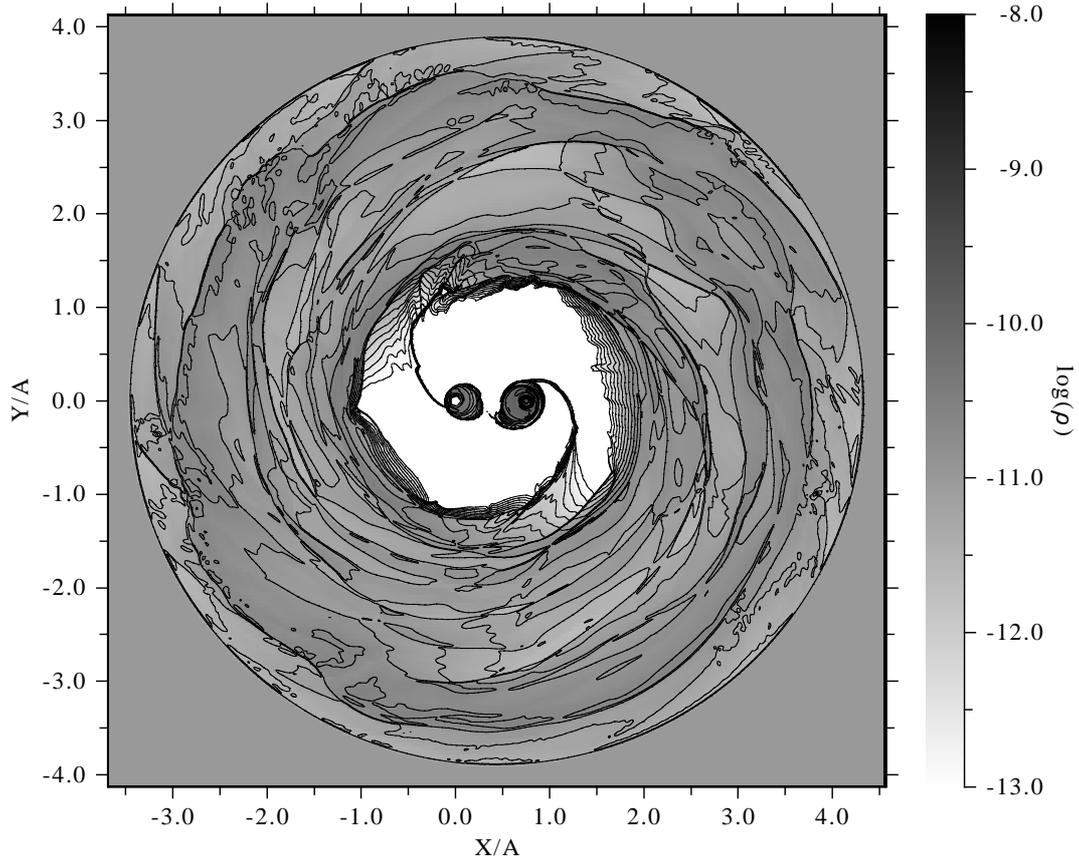, width=14cm}
\end{center}
\caption{ Distribution of the logarithm of the density in the
equatorial plane of the system for the two-dimensional
computations (gray scale and contours).
}\label{xy_big}
\end{figure}

\begin{figure}[ht]
\begin{center}
\epsfig{file=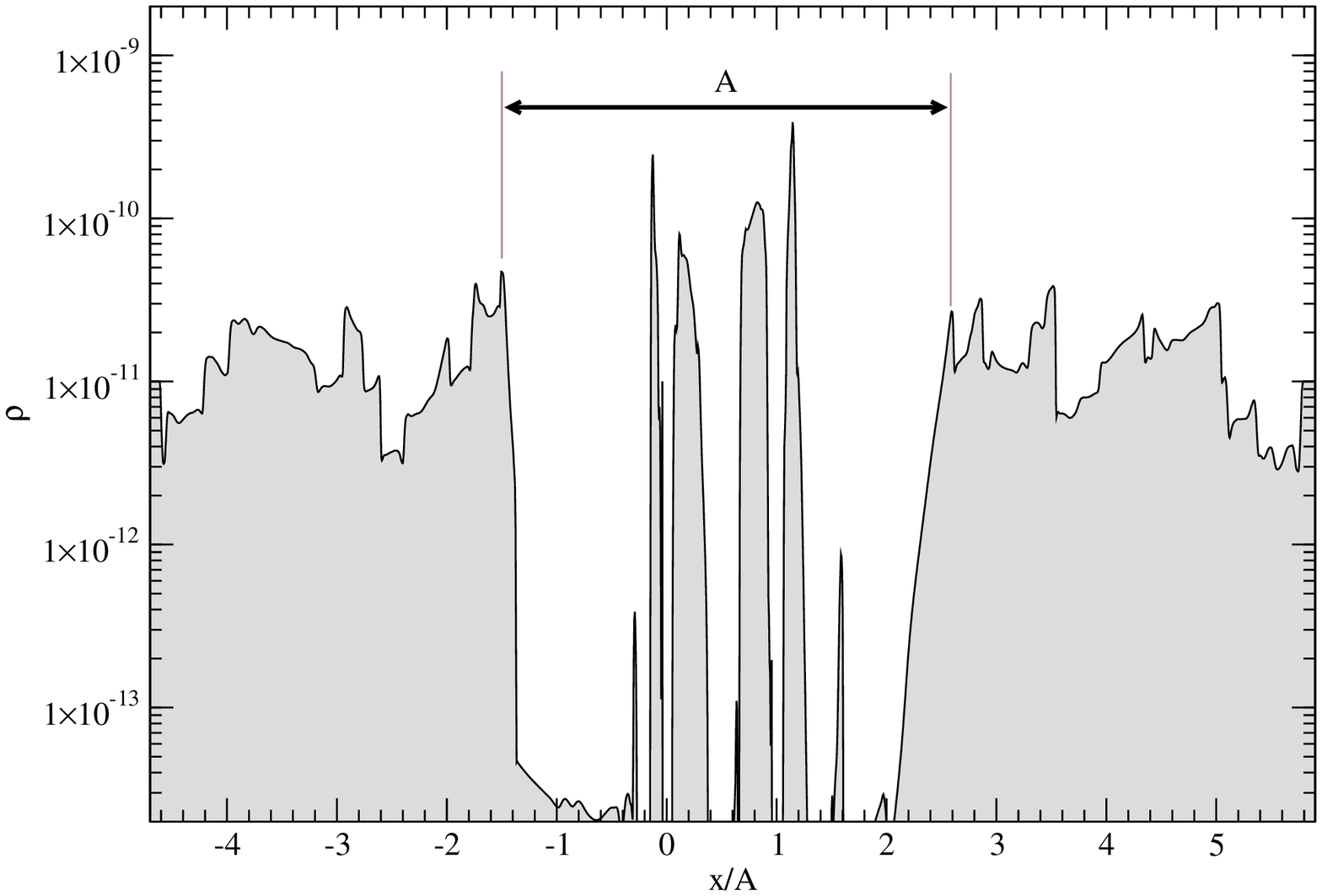, width=14cm}
\end{center}
\caption{ Distribution of density along the line joining the
components. The letter G denotes the gap region.
}\label{rho_x}
\end{figure}

Figure~\ref{xy_big} shows the distribution of the density and
velocity vectors in the equatorial plane of the system, and
Figure~\ref{rho_x} the density distribution along the line joining
the components. The centers of mass of the secondary and primary
components are at the positions (0,0) and (A,0), respectively.
Accretion disks form around the components, bounded by the radii
of the corresponding last stable orbits.Outgoing shocks with the
form of diverging spirals form ahead of the accretion disks (the
rotation of the system is in the counterclockwise direction). At a
distance $R\sim 2\div 3 A$ from the center of mass, the wave
formed by the main component intersects itself, giving rise to a
large number of smaller waves and fragments. The strongly
fragmented wave continues to expand, reaching the edge of the
computational region. The difference in density between the
rarified gap region and the wave is a factor of $\sim 10^3$. The
density difference inside the wave can reach factors of $5\div 7$.
The shape of the rarified region is slightly non-circular: its
extent along the line joining the components is $\sim 20\%$ larger
than in the perpendicular direction.

\begin{figure}[ht]
\begin{center}
\epsfig{file=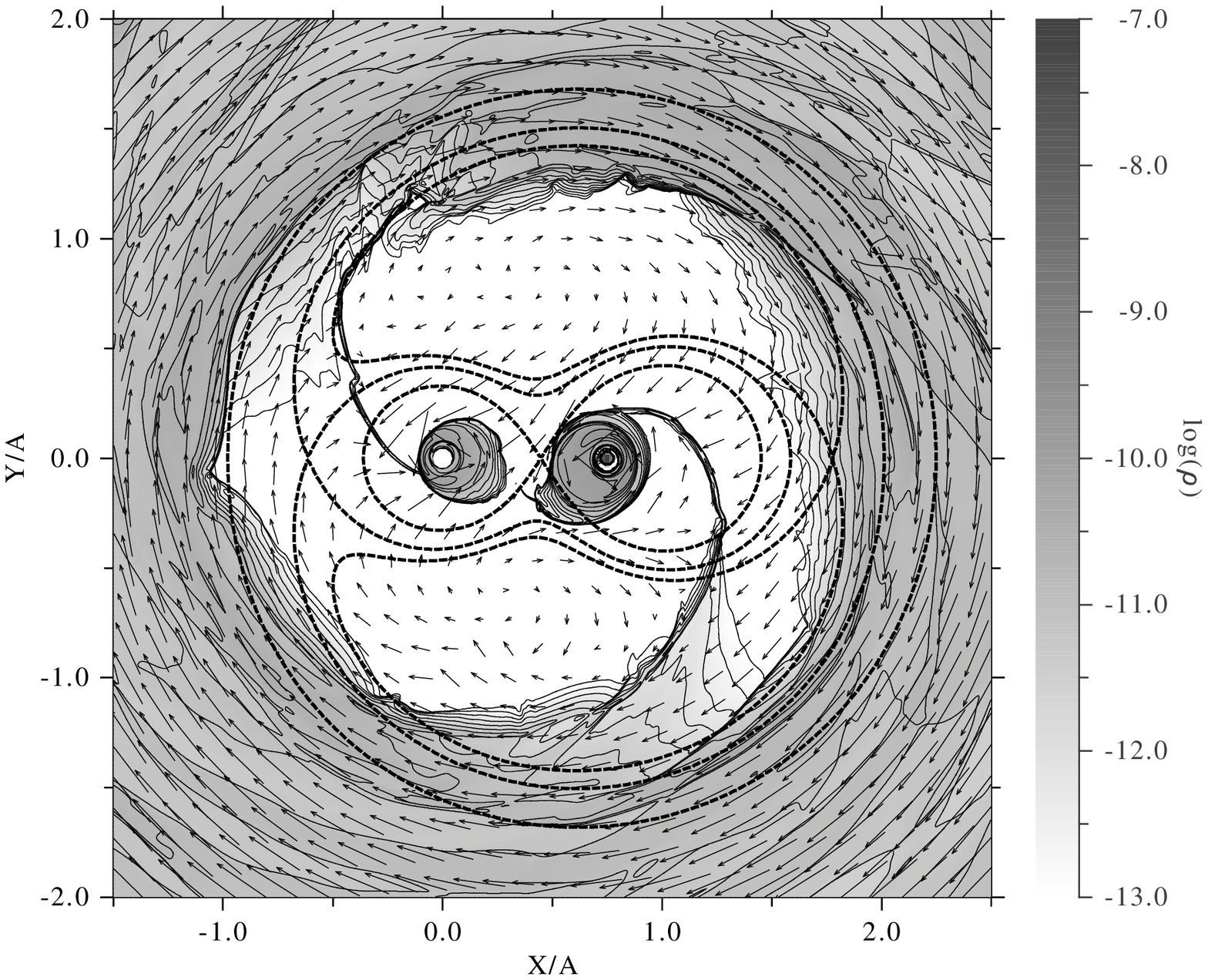, width=14cm}
\end{center}
\caption{ Same as Fig.~\ref{xy_big} for the region near the binary
star. The vectors show the velocity field. Roche equipotential
surfaces are shown by the dashed curves.
}\label{xy_center}
\end{figure}

Figure~\ref{xy_center} shows the region near the binary star on a
larger scale. This clearly shows that the bow shocks of the two
components have slightly different shapes. The bow shock ahead of
the accretion disk of the secondary component is weaker and
displays a lower degree of winding than the wave associated with
the primary. The velocity field in the rarified region indicates a
fairly complex flow pattern.

At some time, the bow shock of the primary component begins to
expand. This shock intersects itself after  $\sim 1.25$ windings
-- the outer edge of the diverging shock collides with the inner
edge of the following winding. Moreover, this shock becomes
perturbed when it collides with the bow shock of the secondary.
These interactions give rise to a large number of shocks that
interfere with each other in the disk.

The bow shock associated with the secondary is also spiral in
form. However, this shock is appreciably weaker, since the
secondary is located in the region of wave rarefication that
arises behind the bow shock of the primary.Moreover, this wave is
disrupted when it collides with the denser shock of the primary
after it has completed less than a fourth of a revolution around
the system.

\begin{figure}[ht]
\begin{center}
\epsfig{file=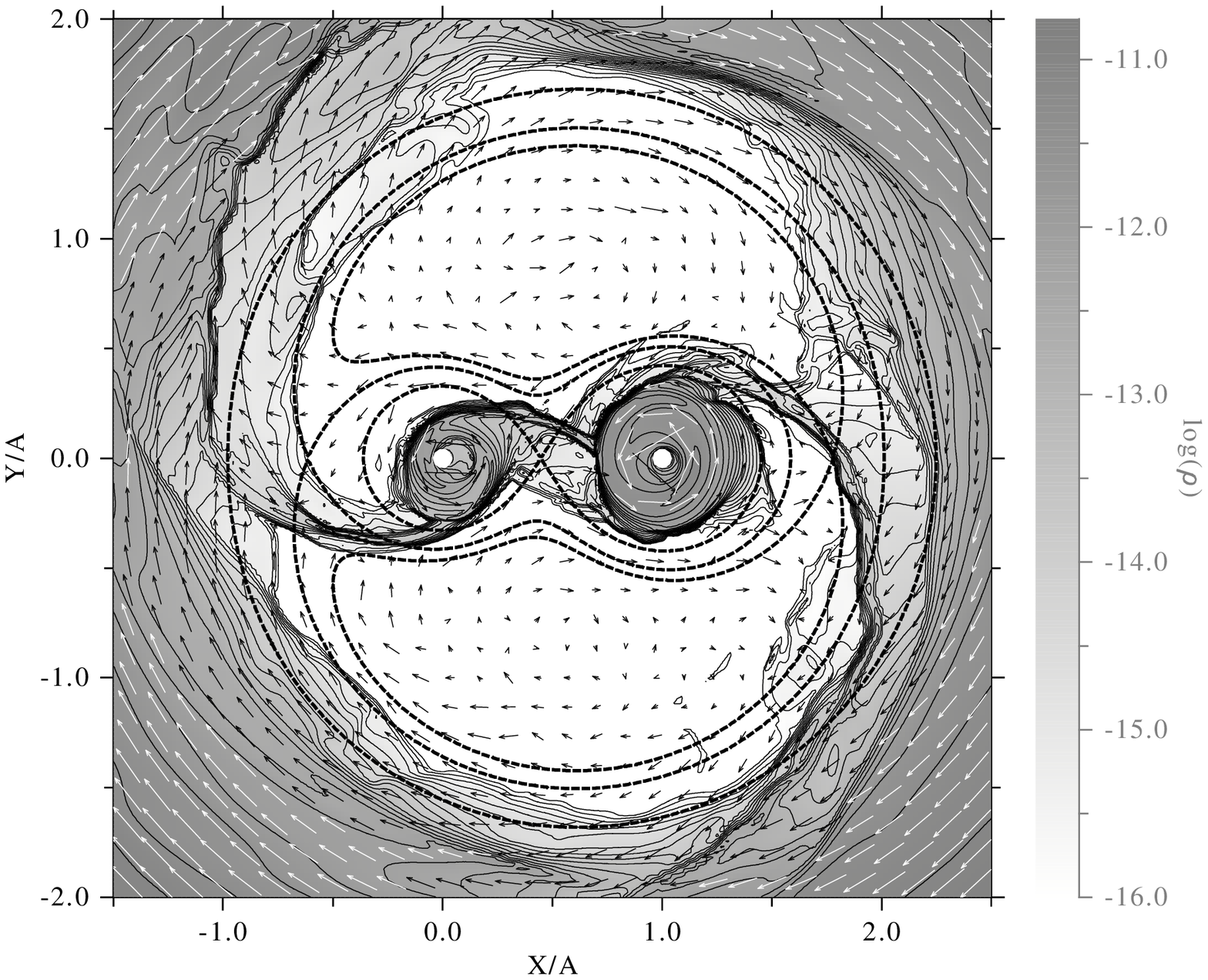, width=14cm}
\end{center}
\caption{ Same as Fig.~\ref{xy_center} for the three-dimensional
computations.
}\label{xy_center_3d}
\end{figure}

Figure~\ref{xy_center_3d} shows the flow structure near the binary
star obtained from the three-dimensional modeling. These results
show all the flow elements obtained in the two-dimensional case,
with their sizes and positions also being similar to those
obtained in the two-dimensional computations.

\section{Conclusions}

Our gas-dynamical solutions show that the gas velocity in the
rarified region near the binary star differs appreciably from a
Keplerian distribution. This suggests that the flow in this region
is determined primarily by gas-dynamical phenomena. The strongest
influence on the flow is exerted by the bow shock that forms due
to the motion of the accretion disk of the more massive component;
accordingly, the radial extent of this wave determines the size of
the inner gap.

The numerical modeling we have considered here shows that the gap
radius is $\sim 2.4 A$ for all the systems studied.
Figure~\ref{modGap} shows the values of the gaps obtained from
observations of various types of stars, together with those
obtained from gas-dynamical modeling. In two cases (stars 1 and
6), the gaps obtained from the modeling are larger than the
corresponding resonance radii; however, the resonance radii are
larger for stars 3 and 4. The best agreement with observations is
obtained for star 5, and the worst agreement for star 2. Finally,
practically all the gap radii obtained from the modeling results
(except for that for star 5) lie below the observed gaps.

Most importantly, we note that stars 3 and 4, whose gaps lie below
the positions corresponding to resonances, have large orbital
eccentricities. The numerical code we used is intended for
modeling systems with circular orbits. To obtain our solutions, we
modeled systems analogous to stars 3 and 4, but with circular
orbits whose radii corresponded to the semi-major axes of these
stars. The SPH modeling carried out in the works~\cite{Artymowicz-Lubow:94,Artymowicz-Lubow:96} shows that the
"cut-off" radius due to the resonances grows when the eccentricity
is increased, and can exceed $2.4A$--the characteristic extent of
the bow shock. Accordingly, the formation of the gap in systems
with high eccentricities may be due to resonances, while the main
role in systems with close to circular orbits is played by the bow
shock.

We also note that the gap sizes derived from observations are
determined in part by the results of modeling the spectra of these
systems. Such modeling usually assumes Keplerian motion of matter
in a protoplanetary disk. However, as the gas-dynamical modeling
shows, the flow is appreciably non-Keplerian, at least in inner
parts of the disk. Moreover, the presence of shocks and dense
rings in the disk must influence the form of the observed spectra.
The lack of allowance for these factors may lead to systematic
errors in the derived gap parameters

\section*{Acnowledgments}

This work was supported by the Basic Research Program of the
Presidium of the Russian Academy of Sciences P-19 "The Origin,
Structure, and Evolution of Objects in the Universe," the Russian
Foundation for Basic Research (project nos. 08-02-00371, 09-
02-00064, 08-02-00928, and 09-02-00993), the federal targeted
program "Science and Science Education Departments of Innovative
Russia in 2009- 2013" (grant of the Ministry of Education and
Science of the Russian Federation "Investigation of Non-stationary
Processes in Stars and the Interstellar Medium at the Institute of
Astronomy of the Russian Academy of Sciences").

%\bibliographystyle{az}
%\bibliography{mrabbrev,diser,my}

\end{document}